\documentclass[aps,prx,preprint,superscriptaddress]{revtex4-1}

\usepackage{graphicx}
\usepackage{chemformula} 
\usepackage{bm}

\newcommand{\hmn}[1]{
  \ensuremath{\begingroup\setupHMN #1\endgroup}%
}

\newcommand{\setupHMN}{%
  \doHMN{-}{\HMNoverline}%
  \doHMN{*}{\HMNminverse}%
  \doHMN{i}{\infty}
}

\newcommand{\doHMN}[2]{%
  \begingroup\lccode`~=`#1
  \lowercase{\endgroup\let~}#2%
  \mathcode`#1="8000
}

\newcommand{\HMNminverse}[1]{\frac{#1}{m}}
\newcommand{\HMNoverline}[1]{\mkern1mu\overline{\mkern-1mu#1\mkern-1mu}\mkern1mu}

\begin{document}


\title{Field-Dependence of Magnetic Disorder in Nanoparticles}



\author{Dominika Z\'{a}kutn\'{a}}
\altaffiliation{present address: Department of Inorganic Chemistry, Faculty of Science, Charles University in Prague, Hlavova 2030/8, 12843 Prague 2, Czech Republic}
\affiliation{Institut Laue-Langevin, 71 Avenue des Martyrs, F-38042 Grenoble, France}
\affiliation{Department f{\"u}r Chemie, Universit{\"a}t zu K{\"o}ln, Greinstr. 4-6, 50939 K{\"o}ln, Germany}
\author{Daniel Ni\v{z}\v{n}ansk\'{y}}
\thanks{deceased}
\affiliation{Department of Inorganic Chemistry, Faculty of Science, Charles University in Prague, Hlavova 2030/8, 12843 Prague 2, Czech Republic}
\author{Lester C. Barnsley}
\altaffiliation{present address: Australian Synchrotron, ANSTO, 800 Blackburn Road, Clayton 3168, Australia}
\author{Earl Babcock}
\author{Zahir Salhi}
\author{Artem Feoktystov}
\affiliation{Forschungszentrum J{\"u}lich GmbH, J{\"u}lich Centre for Neutron Science (JCNS) at Heinz Maier-Leibnitz Zentrum (MLZ), Lichtenbergstr. 1, 85748 Garching, Germany}
\author {Dirk Honecker}
\altaffiliation{present address: Physics and Materials Science Research Unit, University of Luxembourg, 162A Avenue de la Fa\"{\i}encerie, L-1511 Luxembourg, Grand Duchy of Luxembourg}
\affiliation{Institut Laue-Langevin, 71 Avenue des Martyrs, F-38042 Grenoble, France}
\author{Sabrina Disch}
\email{corresponding author. sabrina.disch@uni-koeln.de}
\affiliation{Department f{\"u}r Chemie, Universit{\"a}t zu K{\"o}ln, Greinstr. 4-6, 50939 K{\"o}ln, Germany}


\date{\today}

\begin{abstract}
  The performance characteristics of magnetic nanoparticles towards application,\emph{e.g.} in medicine, imaging, or as sensors, is directly determined by their magnetization relaxation and total magnetic moment. In the commonly assumed picture, nanoparticles have a constant overall magnetic moment originating from the magnetization of the single-domain particle core surrounded by a surface region hosting spin disorder.
  In contrast, this work demonstrates the significant increase of the magnetic moment of ferrite nanoparticles with applied magnetic field.
  At low magnetic field, the homogeneously magnetized particle core initially coincides in size with the structurally coherent grain of 12.8(2) nm diameter, indicating a strong coupling between magnetic and structural disorder. Applied magnetic fields gradually polarize the uncorrelated, disordered surface spins, resulting in a magnetic volume more than 20\% larger than the structurally coherent core. The intraparticle magnetic disorder energy increases sharply towards the defect-rich surface as established by the field-dependence of the magnetization distribution.

  In consequence, these findings illustrate how the nanoparticle magnetization overcomes structural surface disorder.
  This new concept of intraparticle magnetization is deployable to other magnetic nanoparticle systems, where the in-depth knowledge of spin disorder and associated magnetic anisotropies will be decisive for a rational nanomaterials design.
\end{abstract}


\maketitle


\section{Introduction}
The phenomenon of disorder is ubiquitous in structural science, and different qualities of disorder are evident, ranging from the intuitive random disorder to complex types of correlated disorder. Correlated disorder is essential for a large number of functional properties, including polar nanoregions in relaxor ferroelectrics \cite{Xu2008}, colossal magnetoresistance in La$_\mathrm{x}$Ca$_\mathrm{(1-2x)}$MnO$_3$ \cite{Billinge2000}, the entropic disorder in thermoelectrics \cite{Christensen2008}, and correlated spin disorder leading to quasi-particles such as skyrmions \cite{Muhlbauer2009} and magnetic monopoles \cite{Bramwell2009}.
Being intrinsic to nanomaterials, disorder effects such as surface spin disorder\cite{Kodama1996} and surface anisotropy\cite{Garanin2003,Luis2002} in magnetic nanoparticles (NP) crucially determine their magnetization properties including coercivity and superparamagnetism, exchange interactions, and spontaneous magnetization\cite{Skumryev2003,Salazar-Alvarez2008}. These have a pivotal importance for the diverse technological applications of magnetic nanoparticles, such as in recording media\cite{Bader2006}, biomedicine\cite{Pankhurst2003,Tong2017,Yang2018}, catalysis \cite{Niether2018}, or battery materials \cite{Li2015}. The impact of disorder on the heating performance of magnetic nanoparticles has recently been demonstrated \cite{Bender2018,Lak2018,Lappas_PRX_2019}. However, despite the great technological relevance and fundamental importance, the three-dimensional magnetic configuration and the nanoscale distribution of spin disorder within magnetic nanoparticles remains a key challenge.

Surface spin canting or disorder in magnetic NPs is accessible only indirectly using magnetization measurements, ferromagnetic resonance (FMR), M\"{o}ssbauer spectroscopy \cite{Pianciola2015}, X-ray magnetic circular dichroism\cite{Bonanni2018} and electron energy loss spectroscopy\cite{Negi2017}. Spin canting at the NP surface arises from low-coordination sites and a high number of broken exchange bonds of the surface atoms\cite{Masih2012}, and causes a field-dependent shift of the superparamagnetic blocking temperature and exchange bias phenomena\cite{Peddis2008,Toro2017,Kwan2017}. Below the spin glass transition, surface spins freeze in a random configuration\cite{Winkler2005}. In addition, a strong correlation of magnetic and structural disorder is widely accepted\cite{Morales1999,Nedelkoski2017,Bittova2011,Kubickova2014,Wetterskog2013}. In order to reliably discriminate bulk and surface contributions to magnetic disorder, spatial resolution of the intraparticle spin structure is required.

Magnetic small-angle neutron scattering (SANS) is a versatile technique to obtain spatially sensitive information of the spin structure in NPs directly on the relevant nanometer length scale\cite{review2018}. Using half-polarized SANS (SANSPOL), the quantitative magnetization distribution within maghemite NPs has been resolved confirming the presence of spin disorder at the particle surface, but at the same time revealing a significant degree of spin disorder in the entire NP\cite{Disch2012}. Applying SANS with uniaxial polarization analysis (POLARIS) to NP assemblies, a canted magnetic surface shell was reported\cite{Krycka2010,Krycka2014} and the temperature dependence of the spin canting in \ch{CoFe2O4} NPs was derived\cite{Hasz2014}. Micromagnetic simulations of isolated magnetic NPs in a nonmagnetic matrix demonstrated how the interplay between various magnetic interactions leads to nonuniform spin structures in NPs resulting in a strong variation of the magnetic SANS\cite{Vivas2017,Bersweiler2019}. In the context of a polarized SANS study on \ch{Fe3O4}/Mn-ferrite core/shell structures, complementary atomistic magnetic simulations considering a drastically reduced exchange coupling between the core and shell spins revealed no remanence for the shell along with a disordered rather than canted surface spin configuration \cite{Oberdick2018}. Hence, surface spins might potentially be susceptible to intermediate fields, analogous to the spin-flop phase observed in bulk antiferromagnetic oxides \cite{Bhattacharjee2007}. Up to now, all studies of the magnetic nanoparticle spin structure relied on a static picture of a constant, field-independent nanoparticle moment.

\begin{figure}[htbp]
\includegraphics[width=\columnwidth]{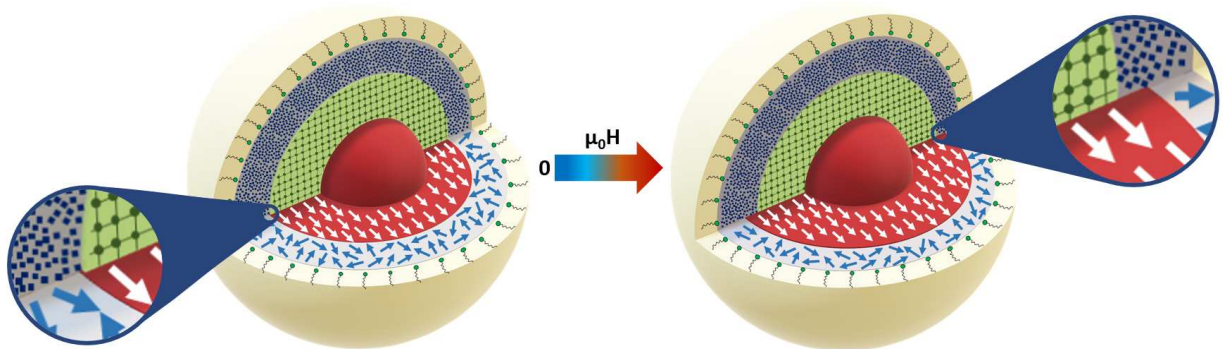}%
\caption{\label{fig:grainsize}Schematic of the structural and field-dependent magnetic NP morphology: the vertical cuts represent the structural morphology, consisting of a structurally coherent grain size (green) and structural disorder (blue) within the inorganic particle (grey). The horizontal cuts represent the magnetic morphology, consisting of a collinear magnetic core (red) and spin disorder (blue) within the inorganic particle surface layer (grey). The particle is surrounded by an oleic acid ligand layer (beige). Structural and magnetic particle sizes are equal in zero field (left), whereas the initially disordered surface spins are gradually polarized in applied magnetic field such that the magnetic radius increases beyond the structurally disordered surface region (right).}
\end{figure}
In this work, we present the field dependence of collinear magnetization and spin disorder in ferrite nanoparticles and derive the associated disorder anisotropy towards the particle surface with spatial resolution. The spontaneous, non-correlated spin disorder at the particle surface is strongly related to structural surface disorder. Remarkably, we observe that with increasing magnetic field the collinear magnetic volume overcomes the structurally coherent particle size. In other words, we demonstrate that the commonly assumed static picture of a constant integral nanoparticle moment with surface spin disorder is not sufficient and needs to be replaced by a field-dependent magnetic nanoparticle core size. This main result of our work is illustrated in \textbf{Figure \ref{fig:grainsize}}. From the field-dependence of the magnetic particle volume, we further extract the spatial extent of spin disorder and derive the associated disorder energy distribution based on a free energy calculation. Consistent with macroscopic magnetization and supported by micromagnetic simulations, our findings demonstrate the intricate nature of \emph{intra}particle disorder anisotropy.

\section{Results and Discussion}
\subsection{NP Structure and Morphology}
\begin{figure}[htbp]
	\begin{center}
		\advance\leftskip-0.4cm
		\advance\rightskip0cm
		\includegraphics[clip,width=14cm]{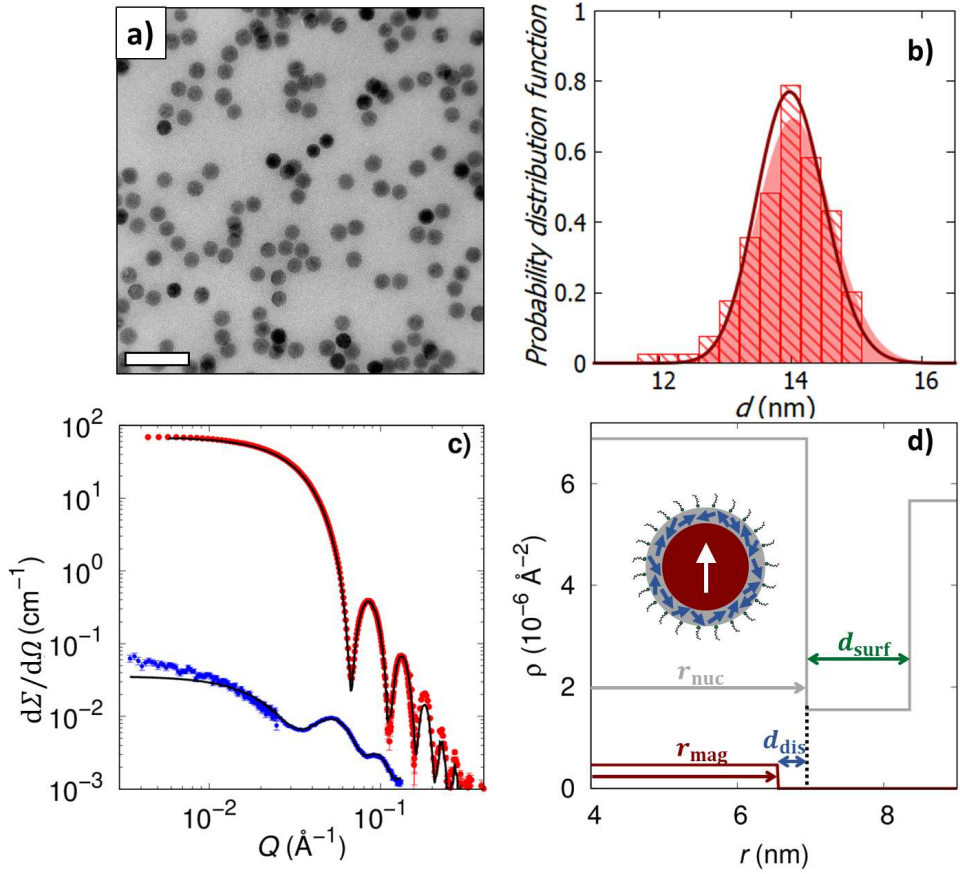}
		\caption{\textbf{a)} TEM bright field micrograph (scale bar: 50\,nm) and \textbf{b)} particle-size histogram based on the evaluation of 200 particles along with log-normal particle size distribution obtained from TEM analysis (red surface) and SAXS refinement (line). \textbf{c)} SAXS (red) and nuclear SANS (blue) data along with form factor fit (black lines) and \textbf{d)} radial profiles of the nuclear ($\rho_\textnormal{n}$, grey) and magnetic scattering length densities ($\rho_\textnormal{mag}$, red). Our model of the magnetic nanoparticle morphology consists of a coherently magnetized particle core with radius $r_\mathrm{mag}$ and a magnetically disordered surface shell with thickness $d_\mathrm{dis}$ within the inorganic NP with radius $r_\mathrm{nuc}$, stabilized by the oleic acid ligand shell with thickness $d_\mathrm{surf}$.}
		\label{fig:NPstructure}
	\end{center}
\end{figure}

The precise evaluation of \textit{intra}particle morphologies such as magnetization distribution and spin disorder optimally requires monodisperse samples of non-interacting magnetic nanoparticles. We therefore synthesized the oleic acid (OA)-capped cobalt ferrite NPs for our study according to Park \textit{et al.}\cite{Park2004} and stabilized them in the non-polar solvent toluene\cite{Zakutna2018}. The sample consists of spherical particles with a log-normal size distribution of 3.1(1)\% and a mean particle radius of $r_\mathrm{nuc}$ = 7.04(5)\,nm as determined using Small-Angle X-ray Scattering (SAXS), which is in excellent agreement with the results obtained from transmission electron microscopy (TEM) analysis (\textbf{Figure \ref{fig:NPstructure}}).
These results define the structural parameters of the inorganic particle core size. A Guinier plateau observed in the lower $Q$ range of the SAXS data further demonstrates the absence of interparticle interactions in toluene dispersion (\textbf{Figure \ref{fig:NPstructure} c)}). The crystal symmetry of the particles determined by powder X-ray diffraction (PXRD) corresponds to the cubic spinel (space group \emph{Fd$\bar{3}$m}) with a lattice parameter of $a$ = 8.362(1)\,\AA, which is slightly smaller than for bulk \ch{CoFe2O4} ($a$ = 8.3919\,\AA), an observation commonly reported for nanosized materials\cite{Frenkel2001}. The determined structurally coherent grain size of $d_\mathrm{XRD}$ = 12.8(2)\,nm (\textbf{App. \ref{sec:PXRD}}) is significantly smaller than the particle size, indicating structural disorder near the particle surface.
An organic ligand shell thickness of $d_\mathrm{surf}$ = 1.4(1)\,nm (\textbf{Figure \ref{fig:NPstructure} d)}) is resolved by the nuclear scattering cross section obtained by SANSPOL. This is reasonable given the theoretical value of fully stretched OA (2.1\,nm) and in good agreement with earlier results on OA-stabilized iron oxide NPs in toluene\cite{Disch2012}. From the X-ray and neutron scattering length densities of the particle core ($\rho_\mathrm{x} = 41.61\cdot10^{-6}$\,\AA$^{-2}$ and $\rho_\mathrm{n} = 6.88\cdot10^{-6}$\,\AA$^{-2}$), a Co cation content of 8.4\,at.-\% is determined according to V\'egard's law\cite{Denton1991}. Assuming neutral overall charge, we consider the formula Co$_y$Fe$_{(8-2y)/3}$O$_4$, where $y$ = 1 represents the cobalt ferrite spinel structure and $y$ = 0 corresponds to maghemite ($\gamma$-\ch{Fe2O3}), and derive a composition of Co$_{0.22}$Fe$_{2.52}$O$_4$. The stoichiometry is based on M\"{o}ssbauer spectroscopy measurements (\textbf{App. \ref{sec:Mossb}}) demonstrating the absence of Fe$^{2+}$ in the compound.
EDX scans further support a chemically homogeneous crystalline particle core. A line scan reveals 10\,at.-\% Co content within the entire particle, whereas an average composition of 9.1\,at.-\% Co is confirmed by TEM EDX mapping (\textbf{Figure \ref{fig:EDX}}), both in excellent agreement with the composition derived by small-angle scattering.

\subsection{Field dependent magnetization distribution}
Using the precise structural particle morphology determined above as a prerequisite, we ascertain the magnetic nanoparticle morphology via the magnetic scattering amplitude of polarized SANS.
We model the magnetic nanoparticle morphology with a homogeneously magnetized particle core with radius $r_\mathrm{mag}$ and a spin disorder shell of thickness $d_\mathrm{dis}$ towards the surface (\textbf{Figure \ref{fig:NPstructure} d)}). The magnetic particle size distribution is taken equal to the nuclear size distribution.
The in-field or longitudinal magnetization component $M_\mathrm{z}(H)$ is directly related with the magnetic scattering length density of the particle core $\rho_\mathrm{mag}$ determined using polarized SANS (\textbf{Equation \eqref{eq:bH}}).
\begin{figure}[btp]
	\begin{center}
		\includegraphics[width=\textwidth]{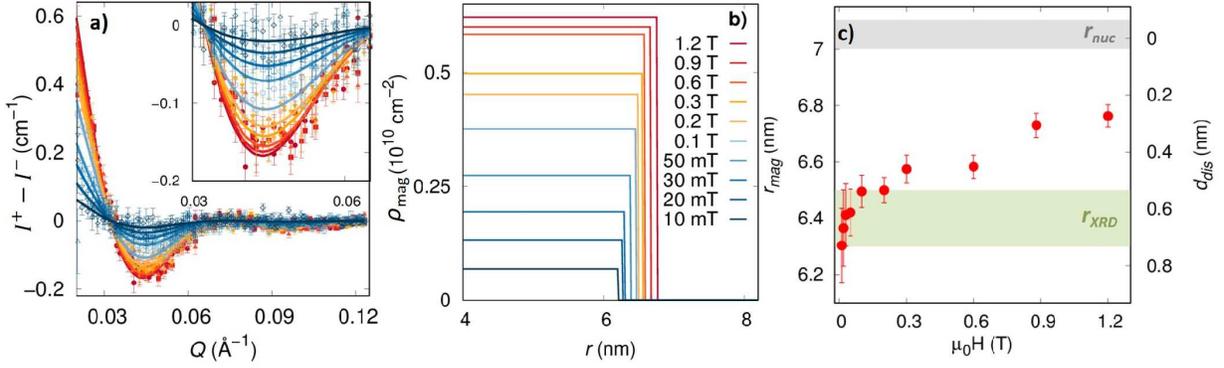}
		\caption{\textbf{a)} Nuclear-magnetic scattering interference term $I^+ - I^-$ (points) at various applied magnetic fields (same color code as in b)) and corresponding fits (lines). Inset: zoomed region of $Q$ = 0.03 - 0.065\,\AA$^{-1}$. \textbf{b)} Field-dependent magnetic scattering length density $\rho_\mathrm{mag}$ profiles. \textbf{c)} Field-dependence of the derived magnetic radius $r_\mathrm{mag}$ and the disordered shell thickness $d_\mathrm{dis}$. The uncertainty intervals of nuclear ($r_\mathrm{nuc}$) and structurally coherent radius ($r_\mathrm{XRD}$) are indicated in grey and green, respectively.
}
		\label{fig:SANSPOLresults}
	\end{center}
\end{figure}
The nuclear-magnetic interference scattering of our sample (\textbf{Figure \ref{fig:SANSPOLresults} a)}) is consistently described only by a field dependent variation of both $\rho_\mathrm{mag}$ and $r_\mathrm{mag}$(\textbf{Figure \ref{fig:SANSPOLresults} b)}) in contrast to a static model using a field-independent $r_\mathrm{mag}$ (\textbf{App. \ref{sec:model}}). The magnetic particle radius $r_\mathrm{mag}(H) < r_\mathrm{nuc}$ increases with magnetic field, starting from $r_\mathrm{mag}$($H_\mathrm{min}$) = 6.3(1)\,nm at the lowest applied magnetic field of $H_\mathrm{min} =$ 11\,mT and attaining $r_\mathrm{mag}$($H_\mathrm{max}$) = 6.76(4)\,nm at the highest applied field of $H_\mathrm{max} =$ 1.2\,T (\textbf{Figure \ref{fig:SANSPOLresults} b,c)}). The spontaneous $r_\mathrm{mag}$($H_\mathrm{min}$) is in excellent agreement with the structurally coherent domain size of 12.8(2)\,nm from PXRD, indicating a structurally homogeneous and spontaneously magnetized particle core smaller than the NP itself.
This observation is in line with reports on reduced magnetic domain size in magnetic NPs, suggested by macroscopic magnetization\cite{Singh2017,Bittova2011,Kubickova2014,Bender2017} and neutron diffraction\cite{Golosovsky2006}. Previous polarized SANS studies indicate spatial correlation of spins near the particle surface giving rise to canted spin structures\cite{Krycka2010,Krycka2014}. Simulations propose a variety of different spin canting scenarios, such as collinear, artichoke, throttled and hedgehog spin structures\cite{Labaye2002,Berger2008}.

\begin{figure}[tbp]
		\includegraphics[angle=0, width=\textwidth]{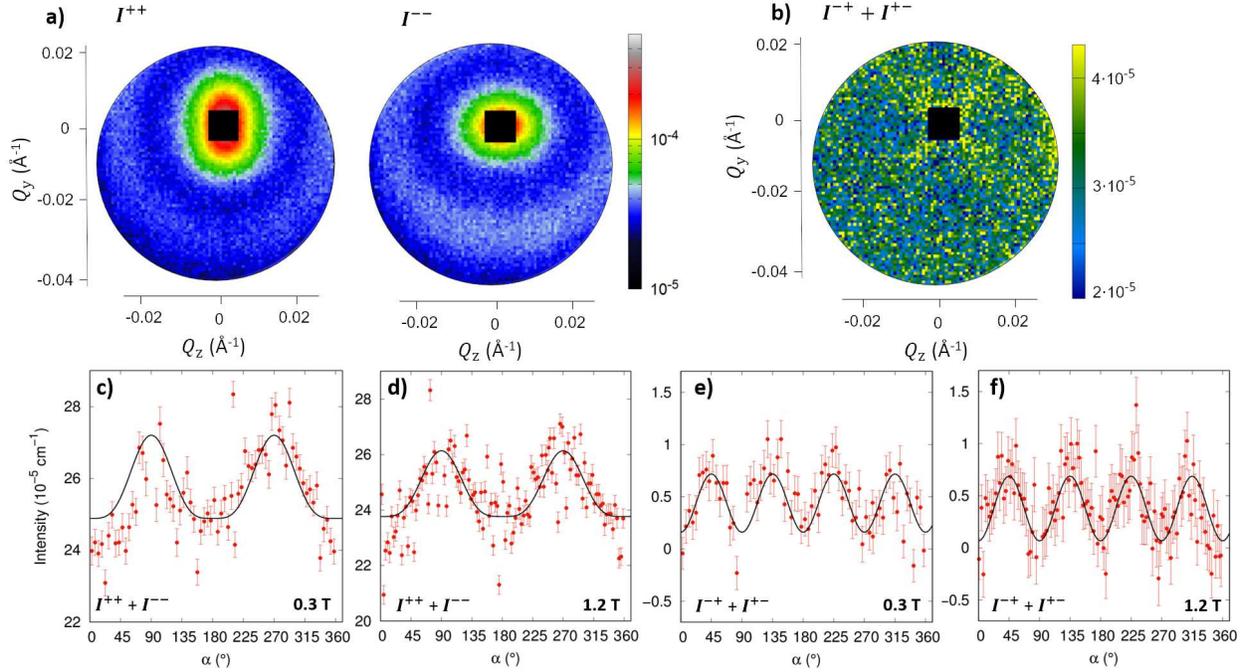}
		\caption{Spin-resolved SANS (POLARIS) by cobalt ferrite nanoparticles: a) Non-spin-flip (NSF, $I^{++},I^{--}$) and b) averaged spin-flip (SF, ($I^{+-}+I^{-+}$)/2) 2D scattering cross sections at an applied magnetic field of 1.2\,T. c-f) NSF and SF azimuthal scattering intensities, radially averaged in a $Q$-range of 0.006 - 0.016\,\AA$^{-1}$, recorded at 0.3 and 1.2\,T with corresponding fits (black line).}
		\label{fig:POLARIS}
\end{figure}

To distinguish between correlated (spin canting) and non-correlated (spin disorder) spins near the NP surface, we performed spin-resolved SANS (POLARIS) on the non-interacting nanoparticles in dispersion (\textbf{Figure \ref{fig:POLARIS}}).
POLARIS gives access to the Fourier transformation of magnetization correlations along the three Cartesian directions 
\cite{review2018}.
In our case of spherical nanoparticles, the transversal magnetization correlations $\lvert\widetilde{M}_\mathrm{\perp}\rvert^2 = \lvert\widetilde{M}_\textnormal{x}\rvert^2 = \mid\widetilde{M}_\textnormal{y}\mid^2$ are assumed to be equal for symmetry reasons.
Based on POLARIS data of two different applied magnetic fields (\textbf{Figure \ref{fig:POLARIS}}), we conclude that the particles do not exhibit a coherently ordered, transversal magnetization component $\lvert\widetilde{M}_\mathrm{\perp}\rvert^2$.
Despite low scattering statistics, in particular in the spin-flip data, the fit parameters of nuclear scattering amplitude, incoherent background, and longitudinal magnetization 
obtained from the different data sets are in excellent, consistent agreement, including the expected slight increase of the longitudinal magnetization following the orientation of the particle moment with applied field (see details in \textbf{App.\ref{sec:POLARIS}}).
The absence of a coherent, elastic scattering contribution originating from transversal magnetization $\lvert\widetilde{M}_\mathrm{\perp}\rvert^2$ is a strong indication of a non-correlated, random spin disorder for our sample, in contrast to the canted spin structures suggested in the literature.

Whereas the existence of surface spin disorder and canting has been under debate in the past, the field-induced reduction of the magnetically disordered surface shell thickness $d_\mathrm{dis}(H) = r_\mathrm{nuc} - r_\mathrm{mag}(H)$ (\textbf{Figure \ref{fig:SANSPOLresults} c)}) revealed in this work is an entirely new observation. At the lowest applied magnetic field of 11 mT, 28(5)\% of the particle volume is associated to a disordered surface with a thickness of $d_\mathrm{dis}$ = 0.7(1)\,nm. The coherently magnetized particle core size, and hence the magnetic particle moment, gradually increases with applied magnetic field, indicating a field-induced alignment of the initially disordered spins even beyond the structurally coherent grain size. At maximum applied field ($\mu_0H_\mathrm{max} =$ 1.2\,T), a non-magnetic surface layer of $d_\mathrm{dis} = 0.28(6)$\,nm\space persists, implying a strong degree of spin disorder in 12(2)\% of the particle volume that cannot be overcome by the magnetic field applied in this study.

\indent The spatially resolved magnetization obtained using SANSPOL gives unprecedented detailed insight into the spontaneous nanoparticle magnetization as valuable complement to standard macroscopic magnetization measurements.
In the conventional picture, the isothermal magnetization for a superparamagnet is described based on the assumption of a field-independent, constant magnetic particle moment. The relative magnetization is described as:
\begin{equation}
\frac{\langle M \rangle}{M_\mathrm{S}} = \langle\cos\gamma(H)\rangle = L(\xi) = \coth\xi - \frac{1}{\xi},
\end{equation}
where $\langle\cos\gamma(H)\rangle$ is the orientation average over the particle ensemble, with the angle $\gamma$ between the magnetic moment of a particle and the applied magnetic field $\mathbf{H}$. The Langevin parameter is given as $\xi$ = $\frac{\mu\mu_0H}{k_\mathrm{B}T}$ with $\mu_0$ the permeability of free space, $\mu$ the integrated particle moment, $k_\mathrm{B}$ denoting the Boltzmann constant and $T$ the temperature.
\begin{figure}[tbp]
	\begin{center}
		\includegraphics[width=10 cm]{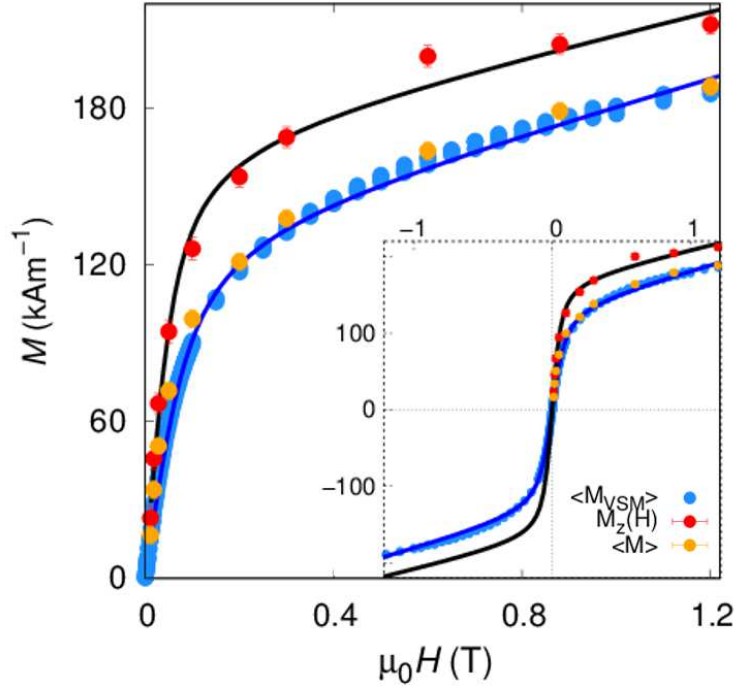}
		\caption{Macroscopic longitudinal magnetization ($\langle M_\mathrm{VSM} \rangle$) measured at room temperature in comparison with the longitudinal particle core magnetization ($M_\mathrm{z}(H)$) and the particle volume averaged magnetization ($\langle M \rangle$), both derived from SANSPOL. Corresponding fits are shown as lines. (inset: full data range for $\langle M_\mathrm{VSM} \rangle$)
}
		\label{fig:VSM}
	\end{center}
\end{figure}
The macroscopic volume magnetization $\langle M_\mathrm{VSM} \rangle$ is typically set in relation with the entire sample (and nanoparticle) volume, \emph{i.e.} disregarding the potentially reduced magnetic volume due to a spin disordered surface region. A Langevin fit of $\langle M_\mathrm{VSM} \rangle$ obtained at 300\,K (\textbf{Figure \ref{fig:VSM})}) yields a particle moment of $\mu = 1.2(1)\cdot10^4 \mu_\mathrm{B}$ with a spontaneous magnetization $M_\mathrm{S,VSM}$ = 135(2) kA/m.
The derived spontaneous magnetization is significantly smaller than for bulk cobalt ferrite (400 kA/m)\cite{Thang2005}.
In addition, an excess paramagnetic susceptibility of $\chi_\mathrm{VSM}= 6.33(6)\cdot10^{-2}$ is evident from the non-saturating magnetization at high applied field. Such excess paramagnetic susceptibility along with reduced spontaneous magnetization as compared to the bulk material is commonly associated to the presence of disordered, misaligned moments \cite{Kodama1996} in addition to the linear high-field susceptibility originating from canted sublattice spins in the bulk material \cite{Clark1968}.
The estimated magnetic particle volume, $V_\mathrm{mag,VSM}$ = $\mu/M_\mathrm{S,VSM} = 8.3(2)\cdot10^{-25}$\,m$^3$ is comparable to the magnetic particle volume $V_\mathrm{mag,SANS} = 1.05(6)\cdot10^{-24}$ m$^3$ derived from the minimal magnetic radius at the lowest applied field. Both magnetic particle volumes are considerably smaller than the morphological NP volume $V_\mathrm{nuc} = \frac{4}{3}\pi r_\mathrm{nuc}^3 = 1.46(3)\cdot10^{-24}$ m$^3$ derived from SAXS. This discrepancy is commonly attributed to surface disorder effects.

The longitudinal magnetization $M_\mathrm{z}(H)$ is based on the coherently magnetized particle core and thus takes into account the variation of the magnetic particle volume. Application of the same Langevin fit as above reveals an enhanced magnetization response (red dots in \textbf{Figure \ref{fig:VSM})}. We extract a spontaneous magnetization $M_\mathrm{S,core}$ = 170(7)\,kA/m, which is larger than $M_\mathrm{S,VSM}$, but still substantially smaller than the bulk saturation magnetization of cobalt ferrite.
The coherently magnetized particle core contributes a particle moment of $\mu = 1.8(2) \cdot10^{4} \mu_\mathrm{B}$ that yields a magnetic particle volume, $V_\mathrm{mag,core}$ = $\mu/M_\mathrm{S,core} = 1.0(1)\cdot10^{-24}$\,m$^3$, in excellent agreement with $V_\mathrm{mag,SANS}$.
We further determine an excess paramagnetic susceptibility of $\chi_\mathrm{core} = 5(1)\cdot10^{-2}$ that is slightly reduced as compared to $\chi_\mathrm{VSM}$.
Our spatially resolved approach thus reveals a homogeneously magnetized particle core with larger magnetization and less spin disorder than expected based on only the macroscopic measurements, but still far from bulk \ch{CoFe2O4} characteristics.

Whereas effects such as spin disorder or sublattice spin canting are commonly parametrized by a linear high-field susceptibility, this simple approach bears the risk to overcompensate further delicate sample-related phenomena such as a bimodal distribution of the particle moment \cite{Bender_SciRep_2017} or the field-dependence of $\mu(\mathrm{H})$ that we observed using polarized SANS.
A closer look into \textbf{Figure \ref{fig:VSM}} reveals signatures of such discrepancies as systematic variations between fit and the experimental data.
Numerical inversion methods for data refinement exist that allow to determine the moment distribution without \emph{a priori} assumptions on a functional form, and hence enable the detection of finer details on the structural and magnetic characteristics of magnetic nanoparticle not retrieved by standard model fits \cite{Berkov2000,VanRijssel2014,Schmidt2017,Bender_SciRep_2017}.
In our case, a model-free analysis of the underlying moment distribution indicates the presence of at least two distinct features which we attribute to the core moments and to the shell magnetization (\textbf{App. \ref{sec:VSM}}).
The actual field-dependence of the particle moment, however, can not be resolved from macroscopic magnetization data alone and requires a spatially sensitive technique such as polarized SANS.

As a consistency proof, we relate the longitudinal magnetization $M_\mathrm{z}(H)$ to the average magnetization of the inorganic particle volume according to $\langle M \rangle = M_\mathrm{z}(H) V_\mathrm{mag}(H)/V_\mathrm{nuc}$ (orange dots in \textbf{Figure \ref{fig:VSM})}). The good agreement with the integral magnetization confirms the reliability of the refinement for a coherently magnetized core with a magnetically disordered surface shell that is further supported by our POLARIS analysis. In consequence, the observed low NP magnetization as compared to the bulk material is a result of both surface spin disorder and reduced magnetization related to a combined effect of the non-stoichiometric amount of Co in the material and structural disorder within the coherently magnetized particle core.
For a composition of Co$_{0.5}$Fe$_{2.5}$O$_4$, a 50\% reduced saturation magnetization compared to nominal \ch{CoFe2O4} has been reported\cite{Torres2015,Nlebedim2012}. For our sample with a composition of Co$_{0.22}$Fe$_{2.52}$O$_4$,
a significant decrease in $M_\mathrm{S}$ may thus be expected.
\begin{figure}[htbp]
	\begin{center}
		\advance\leftskip-0.4cm
		\advance\rightskip0cm
		\includegraphics[clip,width=10cm]{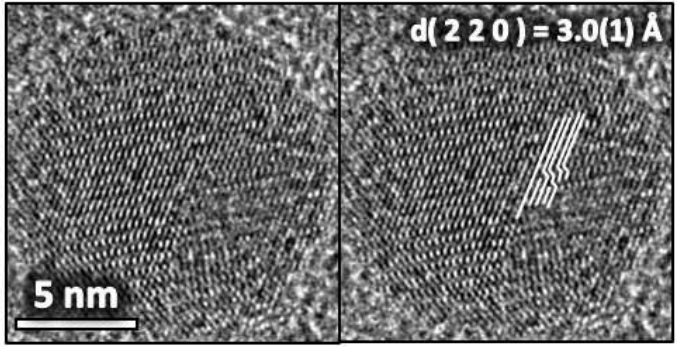}
		\caption{HRTEM micrographs of one representative NP with visible dislocation.}
		\label{fig:HRTEM}
	\end{center}
\end{figure}
In addition, high-resolution TEM (HRTEM) indicates structural disorder in the NP interior including dislocations in the (220) lattice plane (\textbf{Figure \ref{fig:HRTEM}}). Such structural disorder has been observed before in maghemite spinel NPs\cite{Nedelkoski2017,Wetterskog2013}, and is likely correlated with intraparticle spin disorder leading to reduced spontaneous magnetization as well as excess paramagnetic susceptibility in the coherently magnetized NP core.
Detailed investigation of the defected internal structure of small iron oxide nanoparticles has recently revealed uncompensated spin density at atomic-scale interfaces as a result of noncubic local symmetry, in line with enhanced spin canting in the particle interior\cite{Lappas_PRX_2019}.

\subsection{Micromagnetic Approach}
The uncovered field dependence of the magnetic radius may originate in either intrinsic magnetic phenomena, such as surface anisotropy, or structural fluctuations such as gradual lattice distortions near the particle surface. We therefore applied a micromagnetic approach in terms of Ginzburg-Landau theory as introduced by Kronm\"{u}ller and F\"{a}hnle\cite{Kronmuller1980} to describe the magnetic scattering amplitude under the influence of spatially random magnetocrystalline and magnetostrictive fluctuations (\textbf{App. \ref{sec:micromag}}).
The refinement based on a core-shell structure for the magnetic perturbation fields converges for an inner anisotropy constant $K_\mathrm{in}=H_\mathrm{K,in} \cdot M_S=86(52)\, \mathrm{kJ/m^3}$, suggesting a significant amount of magnetic disorder in the particle core interior. Further relevant parameters obtained include an outer anisotropy $K_\mathrm{shell}= 241(91)\, \mathrm{kJ/m^3}$, a shell thickness $d_\mathrm{dis}=1.3(2)\, \mathrm{nm}$, and a spontaneous magnetization of $M_\mathrm{S}=245(19)\, \mathrm{kA/m}$. The derived spontaneous magnetization and shell thickness are in good agreement with the spontaneous magnetization in the particle core and the initial disorder shell thickness determined by SANSPOL. The mean anisotropy field inside the particle $\langle H_\mathrm{K} \rangle=0.6(2)\, \mathrm{T}$ corresponds to a magnetocrystalline anisotropy constant of $\langle K_\mathrm{b} \rangle=149(56)\, \mathrm{kJ/m^3}$, which can be considered as an average value over the entire particle and is in the range of anisotropy constants reported for \ch{CoFe2O4}\cite{Shenker1957}.
This indicates that fluctuations of magnetic parameters, i.e. magnetocrystalline anisotropy and magnetostrictive contributions, are the most likely sources of the variation of the magnetic radius with field. 
In the following, we will consider the magnetic field energy associated to the field dependent variation of the magnetic volume to  extract more detailed, model-independent and spatially resolved information on the extent and strength of the microstructural fluctuations.

\subsection{Spatially resolved disorder energy}
The field-dependent increase of the magnetic volume and the corresponding magnetic field energy occurs in excess of disorder energy that has to be overcome to polarize the initially disordered surface spins (\textbf{App. \ref{sec:Edis}}). The free energy with respect to the initial volume of the magnetic core is given as

\begin{equation}
E_\mathrm{dis}(H) = \mu_\mathrm{0}HM_\mathrm{z}(H)[V_\mathrm{mag}(H) - V_\mathrm{mag}(H_\mathrm{min})].
\end{equation}

\begin{figure}[htbp]
	\begin{center}
		\advance\leftskip-0.4cm
		\advance\rightskip0cm
		\includegraphics[clip,width=12cm]{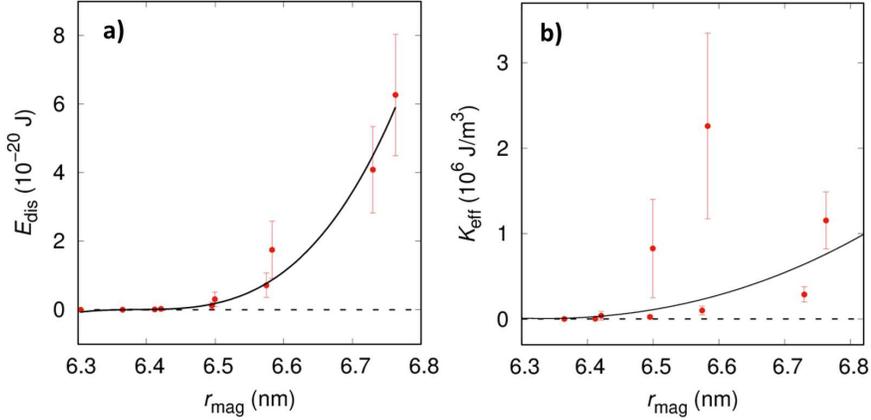}
		\caption{Dependence of \textbf{a)} disorder energy  $E_\mathrm{dis}$ as a function of the coherent magnetic particle radius (black line: spline fit of the data) and \textbf{b)} dependence of the effective disorder energy density $K_\mathrm{eff}$ (black line: derivative of spline in \textbf{a)}).}
		\label{fig:Edis}
	\end{center}
\end{figure}

The gradual growth of the magnetic volume with increasing field is a consequence of a distribution of spin disorder energies such that the spin system is harder to magnetize towards the surface. We attribute this to enhanced structural disorder and significantly reduced exchange interaction near the particle surface. A similarly gradual alignment of surface spins has already been discussed by Kodama, Berkowitz \emph{et al.} \cite{Kodama1996}, who found that surface spins can have multiple meta-stable configurations with the effect that transitions to new equilibrium magnetization states occur with magnetic field.
The magnitude of the surface spin disorder energy shown in \textbf{Figure \ref{fig:Edis} a)} increases up to a value of $E_\mathrm{dis}$(1.2\,T) = 6$\cdot$10$^{-20}$\,J.
Starting from a negligible magnitude in the spontaneously magnetized particle core ($r_\mathrm{mag} <$ 6.3\,nm), the disorder energy density attains a maximum value of $K_\mathrm{eff} = \frac{\partial E_\mathrm{dis}}{\partial V_\mathrm{mag}} \approx 10^6$ Jm$^{-3}$ close to the NP surface (\textbf{Figure \ref{fig:Edis} b)}). We note that the obtained maximum effective energy density value exceeds the bulk magnetocrystalline anisotropy $K_\mathrm{b} = 3.6\cdot10^5$\,Jm$^{-3}$\cite{Shenker1957}.

The derived energy density is understood as the spatially resolved magnetic disorder
anisotropy within the particle. According to phenomenological relations \cite{Yanes2007}, it can be described as a surface anisotropy $K_\mathrm{S} = K_\mathrm{eff}\cdot r_\mathrm{mag}/3$ of the nanoparticle.
Recent particle size dependent studies indicate that surface anisotropy is not necessarily constant\cite{Pisane2017,Singh2017}. Further theoretical studies confirm that the effect of surface anisotropy does not scale with the surface-to-volume ratio, but that surface perturbations penetrate to the NP interior transmitted by exchange interactions leading to a reduced coherent magnetic size\cite{Garanin2003}.
Our approach reveals, for the first time experimentally, how the disorder energy anisotropy varies locally within the particle (\textbf{Figure \ref{fig:Edis} b)}), an aspect that is not accessible by common integral approaches correlating volume averaged values from different batches of NP sizes.
The maximum surface anisotropy of $K_\mathrm{S} \approx 2.3$\,mJm$^{-2}$ is in excellent agreement with N\'eel surface anisotropy\cite{Neel1954} ($0.1 - 1.3$\,mJm$^{-2}$), resulting from broken symmetry at the particle surface and concomitant structural relaxation into the particle core, and it is in the order of magnitude of ferromagnetic materials, \textit{e.g.} Co (1\,mJm$^{-2}$), Er (14\,mJm$^{-2}$), FePt (34\,mJm$^{-2}$), \ch{YCo5} (34\,mJm$^{-2}$)\cite{Berger2008,Gambardella2003}. Ferromagnetic resonance estimates a significantly lower anisotropy for maghemite NPs in ferrofluids (0.03\,mJm$^{-2}$)\cite{Raikher2004} or for non-interacting 7\,nm maghemite NPs  (0.042\,mJm$^{-2}$)\cite{Dormann1996}. However, these values are in good agreement with the volume averaged disorder anisotropy of our sample of $\langle K_\mathrm{S} \rangle $ = 0.096(32)\,mJm$^{-2}$, derived from the maximum disorder energy related to the nuclear particle volume. In this respect, it is noteworthy that the determined values of the surface disorder energies may vary depending on the method and applied magnetic field, as for instance a surface anisotropy of $K_\mathrm{S}$ = 0.027\,mJm$^{-2}$ was obtained from ferromagnetic resonance at 0.1 T\cite{Gazeau1998}.

The gradual decrease of the magnetic disorder parameter (corresponding to enhanced susceptibility) towards the particle interior is likely correlated with reduced structural defect density in the particle core. In addition, spin disorder localized at the particle surface is known to influence the spin configuration in its vicinity via exchange coupling and thus to propagate into the particle interior.
In this respect, hollow spherical maghemite nanoparticles represent interesting model systems to further investigate surface effects on anisotropy and magnetic disorder\cite{Sayed2018}. From magnetization measurements for hollow particles, a strength of the surface anisotropy comparable to the results in this study has been observed\cite{Cabot2009}. Further, based on Monte Carlo simulations, it has been shown that surface spins tend to a disordered  state  due to the competition  between the surface  anisotropy  and  exchange interactions\cite{Cabot2009}.

\section{Conclusion}
This work reveals the field dependence of coherent magnetization and magnetic disorder in highly monodisperse cobalt ferrite nanoparticles and elucidates, for the first time experimentally, the intraparticle disorder energy density with spatial resolution. In contrast to the conventional, static picture, the magnetized core size varies significantly with applied field. This demonstrates that structural surface disorder is overcome by an increasin
g magnetic field in order to gradually polarize the surface spins (\textbf{Figure \ref{fig:grainsize}}).
Indeed, micromagnetic evaluation establishes fluctuations of magnetocrystalline anisotropy and magnetostrictive contributions as the origins of the observed surface spin disorder, and spin-resolved SANS supports non-correlated surface spin disorder rather than spin canting at the particle surface. The spin system is characterized by 12 vol-\% of spin disorder at the particle surface even at a high magnetic field of 1.2\,T. The observed penetration depth of the magnetically perturbed surface region of 0.7\,nm into the nanoparticle interior provides a quantitative insight into the thickness of a magnetic nanoparticle surface.
Our in-depth analysis outperforms the traditional macroscopic characterization by revealing the local magnetization response and by providing quantitative evidence for a spatially varying disorder energy in the nanoparticle, which is not separable from the bulk magnetocrystalline anisotropy by macroscopic characterization methods. The successive increase of the collinear magnetic nanoparticle volume in a magnetic field discloses that one probes the local energy landscape that is constituted of a disorder energy which increases gradually towards the surface. The effective disorder anisotropy increases up to $K_\mathrm{eff} \approx 1\cdot10^{6}$\,Jm$^{-3}$ close to the particle surface, corresponding to a surface anisotropy of $K_\mathrm{S} \approx 2.3$\,mJm$^{-2}$.

\indent The strength of the presented approach is in the unambiguous separation of surface spin disorder from disorder in the nanoparticle core. It can be employed to reliably understand phenomena such as the particle size dependence of the surface disorder and the effects of the chemical environment on the surface spins for varying particle coating. By correlating the magnetic surface disorder energy distribution with structural disorder towards the particle surface, the presented approach furthermore provides indirect insight into the defect concentration and depth profile. Looking beyond magnetic applications, such knowledge of the surface morphology of ferrites plays a decisive role in the diffusion-based fields of heterogeneous catalysis and electrochemistry such as solid-state batteries.

\newpage

\appendix

\section{Synthesis}
Cobalt ferrite NPs were synthesized by thermal decomposition of a mixed Co,Fe-oleate precursor according to Park \textit{et al.}\cite{Park2004}. The oleate precursors were prepared from the respective metal chlorides and freshly prepared sodium oleate as follows: A solution of sodium oleate was prepared by dissolving 66\,mmol (2.64\,g) of NaOH in a mixture of 10\,mL of \ch{H2O} and 20\,mL of EtOH and dropwise addition of 68\,mmol of oleic acid. Water solutions of 15 mL of 8\,mmol (1.9\,g) \ch{CoCl2}$\cdot$6 \ch{H2O} and 16\,mmol (4.32\,g) \ch{FeCl3}$\cdot$6 \ch{H2O} were added to the prepared sodium oleate solution. 60\,mL of hexane and 10\,mL of EtOH were added to the reaction and it was refluxed at 60$^\circ$C for 4 hours. After the reaction cooled down, the oleate complex was washed three times with 50\,mL of water in order to remove NaCl. A brownish viscous mixed oleate complex was obtained by evaporating all solvents including hexane, EtOH and water. In a second step, the ferrite NPs were synthesized by thermal decomposition of 5\,mmol of the prepared oleate precursor with a small amount (1.6\,mL) of additional oleic acid in 25\,mL of octadecene. A heating rate of 2 K/min was applied up to the reflux temperature of 315$^\circ$C which was held for a reflux time of 0.5h. The prepared NPs were precipitated with ethylacetate/EtOH mixture of 1:1 for three times and redispersed in toluene.

\section{Characterization}
\subsection{PXRD \label{sec:PXRD}}
Powder X-ray Diffraction (PXRD) was measured with a PANalytical X'Pert PRO diffractometer equipped with Cu K$_\alpha$ radiation ($\lambda$ = 1.54\,\AA),  a secondary monochromator and a PIXcel detector. The sample was measured in the 2$\theta$  range of 5 - 80$^\circ$ with a step size of 0.03$^\circ$. Rietveld refinement was done in FullProf software\cite{Carvajal1993} using a pseudo-Voigt profile function. The instrumental broadening was determined using a LaB$_6$-reference (SR 660b, NIST).

\begin{figure}[htbp]
	\begin{center}
		\includegraphics[width=0.5\textwidth]{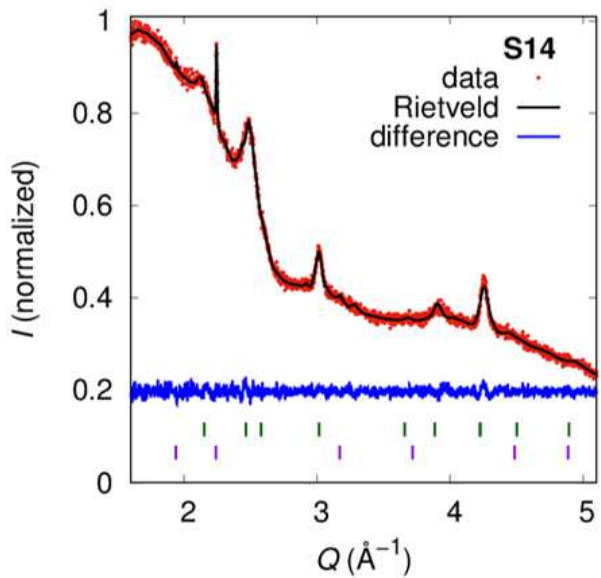}
		\caption{Rietveld refinement of PXRD pattern of cobalt ferrite NPs. The green and purple vertical lines correspond to the Bragg reflections of spinel cobalt ferrite\cite{Natta1929} and of the sodium chloride structure\cite{Hanawalt1938}, respectively.}
		\label{fig:PXRD}
	\end{center}
\end{figure}

The Rietveld refinement of the PXRD pattern shown in \textbf{Fig. \ref{fig:PXRD}}  reveals two phases, a spinel ferrite phase and sodium chloride. The presence of sodium chloride in the sample was due to a non-perfect washing procedure. For the SANS experiment, the preparation was improved by two more purification steps. Nevertheless, this does not affect the structural and the magnetic sample properties.
\begin{table}[htpb]
	\centering
	\begin{tabular}{ccccccc}
		\hline
		\textbf{Parameter} & \ch{CoFe2O4} & & \textbf{Parameter} & & NaCl \\
		\hline
		$a$ (\AA) &  8.362(1) & & $a$ (\AA) & & 5.606(3) \\
		$u$  &  0.234(4) & & & & \\
		O (32 e) &  4.0 & & & & \\
		Co (8 a) &  1.0 & & Na (4 a) & & 1.0 \\
		Fe (16 d) &  2.0 & & Cl (4 b) & & 1.0 \\
		BOV (\AA$^{2}$) & 5.9(1) & & BOV (\AA$^{2}$) & & 64(11)  \\
		\hline
		\textbf{Profile function}& \multicolumn{6}{c}{Thompson-Cox-Hastings pseudo-Voigt}  \\
		\hline
		Y (0.01$^\circ$) & 0.66(4) & & & & 0.037(6)  \\
		zero shift (0.01$^\circ$) &  0.0054(2) & & & & 0.0054(2)\\
		R$_\mathrm{f}$ (\%) &  11.4 & & & & 19.1 \\
		R$_\mathrm{B}$ (\%) &  15.5 & & & & 11.0 \\
		R$_\mathrm{wp}$ (\%) &  2.25 & & & & 2.25 \\
		R$_\mathrm{exp}$ (\%) &  1.69 & & & & 1.69 \\
		$\chi^2$ &  1.77 & & & & 1.77 \\
		\hline
		\textbf{Background function} & \multicolumn{6}{c}{interpolation through 30 points} \\
		\hline
		\textbf{Refined parameters} &  5 & & & & 3 \\
		\hline
		\textbf{Total fit parameters} &  \multicolumn{6}{c}{38} \\
		\hline
	\end{tabular}
	\caption{Rietveld refinement results for the \ch{CoFe2O4} main phase (\hmn{F d -3 m}) with NaCl impurity (\hmn{F m -3 m}), summarizing the lattice parameters $a$, the oxygen site $u$, the occupancy parameters (not refined), the overall isotropic displacement BOV, and the Lorentzian broadening Y.}
	\label{table:PXRD}
\end{table}

\newpage
\subsection{Macroscopic magnetization} \label{sec:VSM}
Vibrating Sample Magnetometry (VSM) measurements were carried out on an ADE EV7 Magnetics Vibrating Sample Magnetometer. 36 µl of the dilute NP dispersion was sealed in a Teflon crucible and placed on a glass sample holder. Room temperature (298 K) magnetization data were collected in a field range $\pm$1.9 T with a head drive frequency of 75 Hz. The diamagnetic contribution of sample holder and solvent was measured independently using a reference measurement of 36 µl of toluene.

\begin{figure}[tbp]
	\begin{center}
		\includegraphics[width=0.9\textwidth]{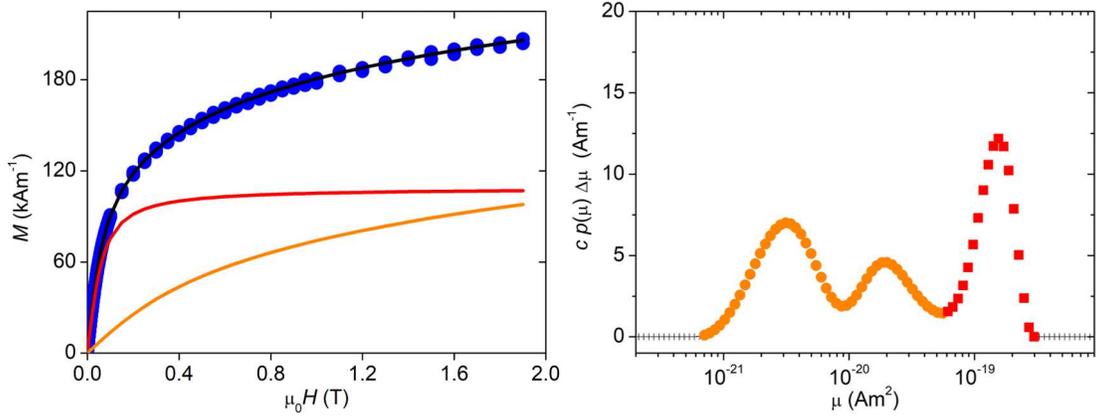}
		\caption{Macroscopic magnetization measurement analyzed by numerical inversion. a) VSM data (blue data) and reconstructed data by numerical inversion (black line). The red and orange contributions arise from the extracted large and small magnetic moment distributions indicated in b). $c$ denotes a scaling factor and $\Delta \mu$ the binning width.}
		\label{fig:VSMnum}
	\end{center}
\end{figure}

Additional to the analysis with a single Langevin term (\textbf{Figure \ref{fig:VSM}}), we also performed an analysis with a distribution $p(\mu)$ of apparent magnetic moments extracted by numerical inversion according to \cite{Bender2017,Bender_Inversion}. The extracted distribution of magnetic moments clearly reveals a central peak responsible for the low field magnetization as well as two features with lower moments assigned to the higher-field susceptibility (\textbf{Figure \ref{fig:VSMnum}}). The central peak is attributed to the integrated nanoparticle moments and is located in the range of $\mu = 1-2 \cdot10^{-19} \mathrm{Am^2}$ with a maximum at $1.55 \cdot10^{-19} \mathrm{Am^2}$, corresponding to $1.7 \cdot10^{4} \mu_\mathrm{B}$. This is in general agreement with our Langevin analysis (\textbf{Figure \ref{fig:VSM}}), but reveals a moment distribution broader than expected due to a distribution in the magnetic nanoparticle volume and potentially variation of the saturation magnetization (\textbf{Figure \ref{fig:VSMnum}b}). The lower moments in the range of $\mu = 10^{-21} - 10^{-19} \mathrm{Am^2}$ are attributed to disordered contributions in the sample.

\subsection{M\"{o}ssbauer spectroscopy \label{sec:Mossb}}
M\"{o}ssbauer spectroscopy of $^{57}$Fe was measured on a Wissel spectrometer using transmission geometry and a proportional detector at ambient temperature without magnetic field. An $\alpha$-Fe foil was used as standard, and spectra fitting was carried out using the Wissel NORMOS routine \cite{NORMOS}.

\begin{figure}[htbp]
	\begin{center}
		\advance\leftskip-0.4cm
		\advance\rightskip0cm
		\includegraphics[clip,width=0.7\textwidth]{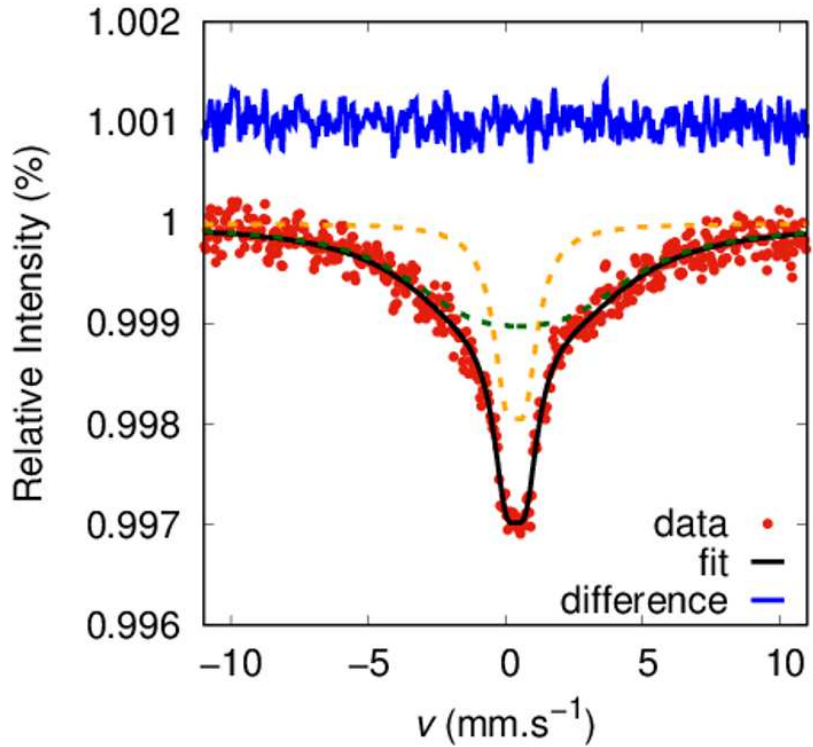}
		\caption{Room temperature M\"{o}ssbauer spectrum. Orange and green dashed lines correspond to the fitted doublet and broad sextet subspectra, respectively.}
		\label{fig:Mossb}
	\end{center}
\end{figure}

\textbf{Figure \ref{fig:Mossb}} presents a room temperature M\"{o}ssbauer spectrogram of the nanoparticles under study. The spectrogram is comparable to M\"{o}ssbauer spectroscopy by maghemite nanoparticles of similar size\cite{rebbouh_PRB_2007} close or above the blocking temperature and was fitted with a singlet and a sextet subspectra including a broad Gaussian distribution due to hyperfine fields or relaxation. We attribute the different subspectra to a distribution of relaxation rates in the nanoparticle sample near the blocking temperature, resulting from the distribution in particle size and, hence, anisotropy energy \cite{Tronc2000}.

The obtained values for the isomer shifts of both subspectra (0.37 and 0.47 mms$^{-1}$ for the singlet and sextet, respectively) clearly indicate the exclusive presence of Fe$^{3+}$ in the sample.

%


\subsection{HRTEM and EDX}
High Resolution Transmission Electron Microscopy (HRTEM) was done on a JEOL JEM 2200FS operated at 200 kV with Schottky emitter using bright field (BF) mode, scanning transmission mode (HRSTEM), energy electron loss spectroscopy (EELS) and energy dispersed (EDS) mapping. The samples were obtained by dropping the toluene dispersion on a coated copper grid.

%
\begin{figure}[htbp]
	\begin{center}
		\advance\leftskip-0.4cm
		\advance\rightskip0cm
		\includegraphics[width=0.6\textwidth]{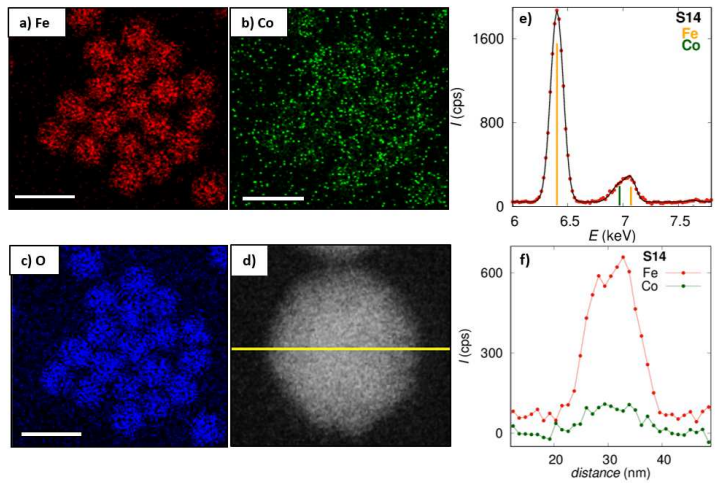}
		\caption{\textbf{a-c)} EDX map of the distribution of Fe, O and Co over a region containing 19-20 particles (Scale bars of EDX maps: 10\,nm) with \textbf{e)} spectra reconstructed from the indicated area. \textbf{f)} EDX line scan over an arbitrarily selected particle, displayed in d).}
		\label{fig:EDX}
	\end{center}
\end{figure}

\newpage

\subsection{SAXS}
Small-angle X-ray scattering (SAXS) measurements were performed at the Gallium Anode Low-Angle X-ray Instrument (GALAXI)\cite{Kentzinger2016} at JCNS, Forschungszentrum J\"{u}lich, Germany. Dilute NP dispersions in toluene ($c$ = 7\,mg/mL) were sealed in quartz capillaries and measured using a wavelength of $\lambda$ = 1.3414\,\AA\space at two detector distances of 853 and 3548\,mm. The data were recorded on a Pilatus 1M detector, radially averaged and normalized to absolute units using FEP 1400\,\AA\space ($d$ = 0.35\,mm) as reference material and toluene background subtraction.

\subsection{SANSPOL} Half-polarized small-angle neutron scattering (SANSPOL) was performed at the D33 instrument\cite{Dewhurst2016} at ILL, Grenoble, France. Dilute NPs dispersions in d$_8$-toluene ($c$ = 7\,mg/mL) were measured at ambient temperature and under applied horizontal magnetic fields up to 1.2\,T. Two instrument configurations were used with detector distances of 2.5\,m and 13.4\,m and collimations of 5.3\,m and 12.8\,m, respectively. The incident neutron beam was polarized using a V-shaped supermirror polarizer. The efficiencies of the flipper and supermirror were 0.98 and 0.94, respectively, at a neutron wavelength of 6\,\AA. Data reduction was performed using the GRASP software\cite{Grasp}.

\begin{figure}[htbp]
	\begin{center}
		\advance\leftskip-0.4cm
		\advance\rightskip0cm
		\includegraphics[width=0.6\textwidth]{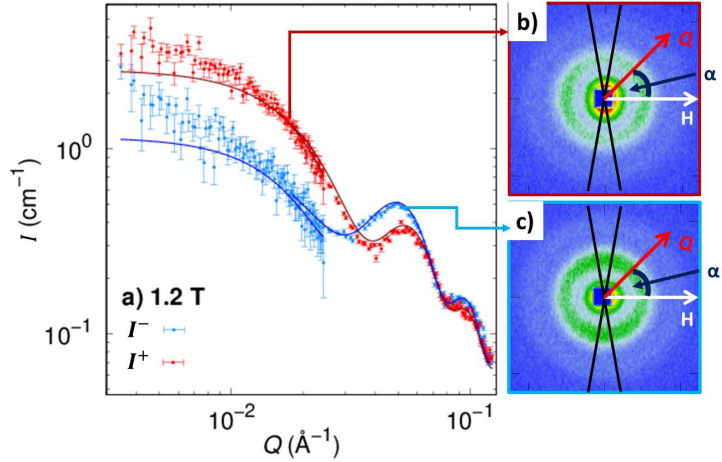}
		\caption{\textbf{a)} SANSPOL scattering cross sections $I^+$ and $I^-$ recorded at 1.2\,T along with refinement (full lines). \textbf{b)} and \textbf{c)} corresponding 2D scattering patterns showing the 20$^\circ$ sectors around an angle of $\alpha$ = 90$^\circ$ between the scattering vector $\textbf{Q}$ and the applied magnetic field $\textbf{H}$.}
		\label{fig:SANSPOL_SI}
	\end{center}
\end{figure}

\indent The SANSPOL cross section of dilute, non-interacting NPs in a magnetic field applied perpendicular to the neutron beam direction is expressed as\cite{Wiedenmann2005,review2018}

\begin{equation}\label{eq:1}
I^{\pm} =  F_\textnormal{N}^{2}(\textnormal{Q}) \mp 2F_\textnormal{N}(\textnormal{Q})F_\textnormal{M}(\textnormal{Q})L(\xi)\sin^2\alpha \\
+ F_\textnormal{M}^{2}\left[\frac{2L(\xi)}{\xi} - \sin^2\alpha\left( \frac{3L(\xi)}{\xi} -1 \right) \right],
\end{equation}

with the azimuthal angle $\alpha$ between the applied magnetic field $\mathbf{H}$ and scattering vector $\mathbf{Q}$, the Langevin function $L(\xi)$ with  $\xi = \mu\mu_0H/\textnormal{k}_\textnormal{B}T$, where $\textnormal{k}_\textnormal{B}$ is the Boltzmann constant, $T$ the temperature, $\mu_0$ the permeability of the free space and $\mu$ the integrated particle moment. According to equation \eqref{eq:1}, the purely nuclear scattering contribution $F_\mathrm{N}^2(Q)$ is accessible from the 2D SANSPOL pattern for $\textbf{Q} \parallel \textbf{H}$ ($\sin^2\alpha = 0$) and a saturating magnetic field ($L(\xi)/\xi = 0$). The longitudinal magnetic scattering amplitude $F_\mathrm{M}(Q)$ is accessible via the nuclear-magnetic interference term $I^+ - I^- = -4F_\mathrm{N}(Q)F_\mathrm{M}(Q)L(\xi)\sin^2\alpha$.
One can assume that the particle is in a single domain state for all fields except for a surface region with reduced magnetization, i.e. the magnetization state of the particle core does not change with field. The integral magnetization is described by Langevin behavior corresponding to the reorientation of the particle moment along the field direction. The magnetic particle moment, increasing with magnetic field, is given by $\mu(H)= F_M(Q=0,H)/b_H = V(H) \cdot \rho_{\mathrm{mag}}/b_H$ and is used as input value for the Langevin function $ L(\xi)$.
The strength of the magnetic scattering is proportional to the magnetic scattering length density $\rho_{\mathrm{mag}}$ that is related to the effective longitudinal magnetization component  $M_z(H)$ of the core according to:

\begin{equation}\label{eq:bH}
\rho_{\mathrm{mag}}=b_H \cdot M_z(H)= b_H \cdot M_s \cdot L(\xi)\, ,
\end{equation}

where $b_\mathrm{H} = 2.91 \cdot 10^8\,\mathrm{A}^{-1} \mathrm{m}^{-1}$ is the magnetic scattering length. The difference method ($I^+ - I^-$) has the advantage that it eliminates background scattering contributions such as incoherent scattering, potential non-magnetic contaminations in the sample, or spin-misalignment contributions arising from moments deviating randomly from the field axis. Complementary refinements of the individual $I^+$ and $I^-$ cross sections (\textbf{Fig. \ref{fig:SANSPOL_SI}}) confirm consistency of the results.

\begin{table}[htpb]
	\centering
	\begin{tabular}{ccccccc}
		\hline
		\textbf{Field}&$\bm{r}_\mathbf{mag}$ &$\bm{d}_\mathbf{dis}$ & $\bm{V}_\mathbf{mag}$ & $\bm{\rho}_\mathbf{mag}$ &$\bm{M}_\mathbf{z}$  &$\bm{\langle M \rangle}$\\
		 (T) & (nm) & (nm) & (nm$^{3}$) & ($10^{-7}$\,\AA$^{-2}$) & (kA/m) & (kA/m) \\
		\hline
		0.011 & 6.30(13) & 0.74(14) & 1047(65) & 0.67(7) & 23(2) & 16(2)\\
		0.018 & 6.36(14) & 0.68(15) & 1078(71) & 1.33(6) & 46(2) & 34(3)\\
		0.028 & 6.41(11) & 0.63(12) & 1103(57) & 1.94(7) & 67(2) & 50(3)\\
		0.05 &  6.42(8) & 0.62(9) & 1108(41) & 2.74(7) & 94(2) & 71(4)\\
		0.1 & 6.50(6) & 0.55(8) & 1148(32) & 3.67(7) & 126(2) & 99(4)\\
		0.2 & 6.49(5) & 0.55(7) & 1145(26) & 4.48(7) & 154(2) & 121(4)\\
		0.3 & 6.57(5) & 0.47(7) & 1188(27) & 4.92(8) & 169(3) & 137(5)\\
		0.6 & 6.58(4) & 0.46(6) & 1193(22) & 5.82(7) & 200(2) & 163(5)\\
		0.88 & 6.73(4) & 0.31(6) & 1277(23) & 5.95(8) & 204(3) & 179(5)\\
		1.2 & 6.76(4)  & 0.28(6) & 1294(23) & 6.17(7) & 212(2) & 188(6)\\
		\hline
	\end{tabular}
	\caption{Parameters refined from field-dependent SANSPOL data, with the magnetic particle radius $r_\mathrm{mag}$, the disorder shell thickness $d_\mathrm{dis}$ and the magnetic scattering length density  SLD$_\mathrm{mag}$. Derived parameters include the magnetic particle volume $V_\mathrm{mag}$ and longitudinal magnetization $M_\mathrm{z}$ as well as average particle magnetization $\langle M \rangle$, considering a nuclear particle volume of $V_\mathrm{nuc}$ = 1462(31)\,nm$^{-3}$ with $r_\mathrm{nuc}$ = 7.04(5)\,nm.}
	\label{table:SANSPOL}
\end{table}

\newpage

\subsection{NP magnetic morphology: static vs. field-dependent magnetic particle volume}\label{sec:model}
In order to prove the validity of our non-static, field-dependent model of the magnetic form factor, we compare it here with a SANSPOL evaluation based on the commonly used static, field-independent magnetic morphology. In this case, also a core-shell model consisting of a collinearly magnetized particle core and a disordered surface shell is considered. The magnetic core size $r_\mathrm{mag}$ is refined in the highest field data (for its best statistics in the nuclear-magnetic interference term) and held constant for all other field-dependent SANS data, leaving the magnetic scattering length density $\rho_\mathrm{mag}$ as the only field-dependent fit parameter.
\begin{figure}[htbp]
	\begin{center}
		\includegraphics[width=\textwidth]{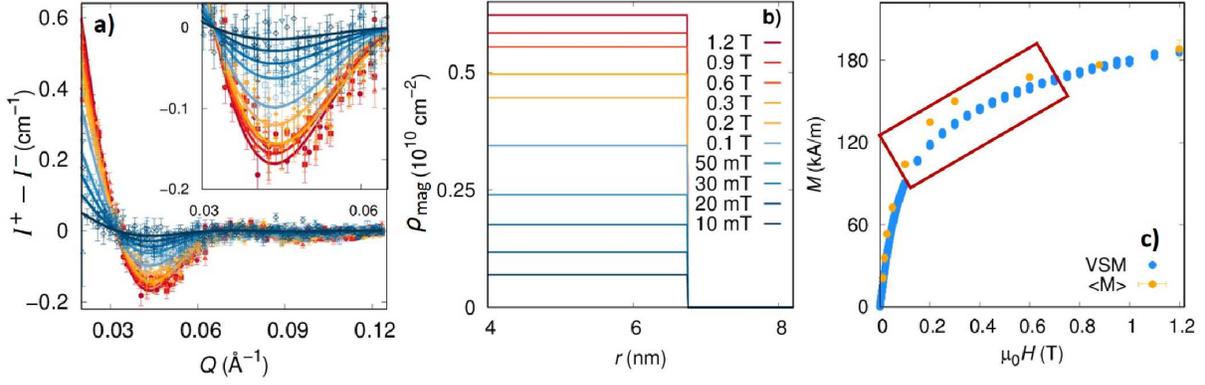}
		\caption{\textbf{a)} Nuclear-magnetic scattering interference term $I^+ - I^-$ (points) at various applied magnetic fields (same color code as in b)) and corresponding fits (lines). Inset: zoomed region of $Q$ = 0.03 - 0.065\,\AA$^{-1}$. \textbf{b)} Field-dependent magnetic scattering length density $\rho_\mathrm{mag}$ profiles. \textbf{c)} Macroscopic longitudinal magnetization $\langle M_\mathrm{VSM} \rangle$ measured at room temperature in comparison with the particle volume averaged magnetization $\langle M \rangle$ as derived from SANSPOL. Deviation in the intermediate field range is indicated (red box).
}
		\label{fig:SANSPOLstatic}
	\end{center}
\end{figure}
\begin{table}[htpb]
	\centering
	\begin{tabular}{ccc}
		\hline
		\textbf{Field}&$\bm{\chi}^2_\mathbf{red}$ &$\bm{\chi}^2_\mathbf{red}$ \\
		 (T) & (static) & (field-dependent) \\
		\hline
		1.2   & 1.155 & 1.155 \\
		0.88  & 1.356 & 1.341 \\
		0.6   & 1.194 & 1.075 \\
		0.3   & 1.433 & 1.38 \\
		0.2   & 1.121 & 1.088 \\
		0.1   & 1.053 & 1.004 \\
		0.05  & 1.043 & 1.019 \\
		0.028 & 1.015 & 1.007 \\
		0.018 & 0.669 & 0.663 \\
		0.011 & 0.801 & 0.817 \\
		\hline
	\end{tabular}
	\caption{Reduced $\chi^2$ values for refinement of the field-dependent SANSPOL data according to either a static or a field-dependent magnetic morphology model. Fit results are shown in \textbf{Fig. \ref{fig:SANSPOLresults}} and \textbf{Fig. \ref{fig:SANSPOLstatic}}, respectively.}
	\label{table:SANSPOL_Chi2}
\end{table}

Results of the static model are presented in \textbf{Fig. \ref{fig:SANSPOLstatic}}, and \textbf{Tab. \ref{table:SANSPOL_Chi2}} provides a direct comparison of the obtained reduced $\chi^2$ for both models. We note that throughout all data sets, the obtained reduced $\chi^2$ is improved for the field-dependent $r_\mathrm{mag}$  model by a few percent. In the very low field, the reduced $\chi^2$ below unity for both models indicates that the fits are overrated. We attribute this to the very small magnetic scattering signal in comparison to the measurement uncertainty at such low applied field. However, the field-dependent $r_\mathrm{mag}$ model yields a better fit of about 3\% on average (1.05 as compared to 1.08 for the static model), and of 5 to 10 \% for intermediate fields (0.1-0.6 T). This indicates that the model with variable $r_\mathrm{mag}$ improves the fit significantly.

The main effect is directly visible in the comparison to the macroscopic magnetization, where the SANSPOL result $\langle M \rangle$ deviates strongly from the VSM data $\langle M_\mathrm{VSM} \rangle$ as indicated by the red box in \textbf{Fig. \ref{fig:SANSPOLstatic} c}.
Comparison of microscopic SANSPOL results with the independently measured macroscopic magnetization is an important proof of consistency.
The deviation shown in \textbf{Fig. \ref{fig:SANSPOLstatic} c} is a clear indication that the applied static model is not sufficient to describe the SANSPOL data reliably. In contrast, the field-dependent model yields excellent agreement of microscopic and macroscopic magnetization as shown in \textbf{Fig. \ref{fig:VSM}}.

In consequence, a consistent analysis of our SANSPOL data, in agreement with macroscopic magnetization measurements, is achieved only by consideration of a field-dependent $r_\mathrm{mag}$. This underlines the need for the spatial sensitivity of magnetic SANS in addition to macroscopic techniques to describe the structural and magnetic details.

\subsection{POLARIS}\label{sec:POLARIS}
Full polarized  small-angle neutron scattering (POLARIS) was done at the KWS-1 instrument\cite{Feoktystov2015} operated by J\"{u}lich Centre for Neutron Science (JCNS) at Heinz Maier-Leibnitz Zentrum (MLZ), Garching, Germany. A dilute non-interacting NP dispersion in d$_8$-toluene was measured at ambient temperature and under applied horizontal magnetic fields up to 1.2\,T. Measurements were performed at the detector distance of 8\,m with a collimation of 8\,m. The incident neutron beam (of 5\,\AA\space neutron wavelength) was polarized using a supermirror polarizer and the polarization of the scattered neutrons was analyzed using a polarized $^{3}$He spin filter cell.
The incident super mirror gave 0.905 for the wavelength of the experiment with a 0.998 flipper efficiency. The incident beam polarization in this case was slightly reduced by a beam depolarization which was later determined to come from the sample slits. At this time off-line polarized $^{3}$He cells were used for KWS-1, therefore two different cells named Jimmy and Willy with 8.9 and 10.8 bar cm of $^{3}$He, respectively, were used \cite{Zahir2014}. Both cells provided 100 ($\pm$ 4) hours on beam lifetimes. Jimmy and Willy gave starting/ending unpolarized neutron transmissions of about 0.21 down to 0.17 and  0.20 going down to 0.14 after a typical half day of use corresponding to initial to final polarization analyzing powers of 0.984 down to 0.976 and 0.995 down to 0.992 for each cell, respectively \cite{Feoktystov2015}. Four cell exchanges between the two cells were made during the course of the experiment to maintain good transmission performance.
Data reduction and spin-leakage corrections due to polarization inefficiencies as well as solvent subtraction were performed using qtiKWS software \cite{qtiKWS}, and extraction of the azimuthal intensities was carried out using GRASP software\cite{Grasp}.

The neutron spin resolved non-spin-flip ($I^{\pm\pm}$) and spin-flip ($I^{\pm\mp}$) cross section of dilute, non-interacting NPs in dispersion are expressed as\cite{review2018}:

\begin{equation}\label{eq:4a}
I^{\pm\pm} \propto F_\textnormal{N}^{2} + \lvert\widetilde{M}_\textnormal{y}\rvert^{2}\sin^2\alpha\cos^2\alpha \mp F_N \widetilde{M}_z \sin^2\alpha + \lvert\widetilde{M}_\textnormal{z}\rvert^{2}\sin^4\alpha+bgr,
\end{equation}

\begin{equation}\label{eq:5a}
I^{\pm\mp} \propto \lvert\widetilde{M}_\textnormal{x}\rvert^{2} + \lvert\widetilde{M}_\textnormal{y}\rvert^{2}\cos^4\alpha + \lvert\widetilde{M}_\textnormal{z}\rvert^{2}\sin^2\alpha\cos^2\alpha +bgr,
\end{equation}

with $\lvert\widetilde{M}_\textnormal{x}\rvert^2, \lvert\widetilde{M}_\textnormal{y}\rvert^2$ and $\lvert\widetilde{M}_\textnormal{z}\rvert^2$ the Fourier transforms of the magnetic correlations along the three Cartesian directions. Terms linear in the transversal component $\widetilde{M_y}$ average out in dispersion since the spin distribution can be assumed to be symmetric around the field direction. Furthermore, due to the particle symmetry, we assume the squared Fourier coefficients of the transversal magnetization to be equal ($\widetilde{M}_x^2=\widetilde{M}_y^2=\widetilde{M}_{\perp}^2$).
%
The  non-spin-flip ($I^{\pm\pm}$) and spin-flip ($I^{\pm\mp}$) cross section of dilute, non-interacting NPs in dispersion under an applied magnetic field perpendicular to the neutron beam direction are hence expressed as\cite{review2018}:

\begin{equation}\label{eq:4}
I^{\pm\pm} \propto F_\textnormal{N}^{2} + \lvert\widetilde{M}_\mathrm{\perp}\rvert^{2}\sin^2\alpha\cos^2\alpha \mp F_N \widetilde{M}_z \sin^2\alpha + \lvert\widetilde{M}_\textnormal{z}\rvert^{2}\sin^4\alpha + bgr,
\end{equation}

\begin{equation}\label{eq:5}
I^{\pm\mp} \propto \lvert\widetilde{M}_\mathrm{\perp}\rvert^{2} (1 + \cos^4\alpha) + \lvert\widetilde{M}_\textnormal{z}\rvert^{2}\sin^2\alpha\cos^2\alpha +bgr,
\end{equation}

with $\lvert\widetilde{M}_\mathrm{\perp}\rvert^2$ and $\lvert\widetilde{M}_\textnormal{z}\rvert^2$ the Fourier transforms of the transversal and longitudinal magnetization correlations, respectively, and $bgr$ a scattering background term originating mainly in spin-incoherent scattering contributions.


The spin-flip (SF) and non-spin-flip (NSF) scattering cross sections shown in \textbf{Fig. \ref{fig:POLARIS}a,b} were radially averaged in the $Q$-range of 0.006 - 0.016\,\AA$^{-1}$.
The resulting azimuthal SF (($I^{+-}+I^{-+}$)/2) and NSF (($I^{++}+I^{--}$)/2) cross sections at the measured magnetic field of 1.2 and 0.3\,T are shown in \textbf{Fig. \ref{fig:POLARIS}c-f}.
%
%
The azimuthal SF and NSF intensities show clearly the $\sin^2\alpha\cos^2\alpha$ and $\sin^4\alpha$ anisotropies, respectively, proportional to the longitudinal magnetization $\lvert\widetilde{M}_\textnormal{z}\rvert^2$.
No sign for a $1+\cos^4\alpha$ behavior arising from transversal magnetization correlations is found in the SF data (\textbf{Fig. \ref{fig:POLARIS}e,f}), and also the NSF data can be described without the need for a transversal magnetization component (\textbf{Fig. \ref{fig:POLARIS}c,d}).
%
From SF scattering, the $\lvert\widetilde{M}_\textnormal{z}\rvert^2$ values of 1.5(1) and 1.4(2) cm$^{-1}$ for applied magnetic field of 1.2\,T and 0.3\,T, respectively were obtained. Small background values of 0.43(5) and 0.49(7) cm$^{-1}$ were obtained from the fit and are attributed to the spin incoherent scattering from oleic acid at the nanoparticle surface. The obtained $\lvert\widetilde{M}_\textnormal{z}\rvert^2$ values of 1.59(7) and 1.53(1) cm$^{-1}$ from NSF scattering at 1.2 and 0.3\,T are in great agreement with the received values of $\lvert\widetilde{M}_\textnormal{z}\rvert^2$ values from SF scattering. Background values of 4.82(2) and 4.95(2) cm$^{-1}$ at 1.2 and 0.3\,T, respectively, in the NSF scattering contribution are assigned to the sum of the spin incoherent and nuclear scattering.


\section{Micromagnetic Theory of an Inhomogeneously Magnetized Particle \label{sec:micromag}}
Based on micromagnetic theory\cite{Kronmuller1980}, we can derive an analytical expression for the magnetic scattering amplitude under the influence of spatially random magnetocrystalline and magnetostrictive fluctuations

\begin{equation}\label{eq:micromag}
 F_M(Q,H)= b_H (\mu_0 M_S F_\mathrm{sphere}(Q, r_\mathrm{nuc})-g_H(Q) p)\, .
\end{equation}

The field dependence enters with the dimensionless micromagnetic response function $p= M_S/ (H_\mathrm{eff}(Q,H)+2 \langle H_K \rangle)$ with $\langle H_K \rangle$ the (field-independent) mean magnetocrystalline anisotropy field averaged over the inorganic particle volume. The effective field $H_\mathrm{eff}(H,Q)=H (1+l_H^2 Q^2)$ depends on the applied field H and on the exchange length of the field $l_H(H)=\sqrt{2 A/(\mu_0 M_S H)}$ with the parameter $A$ denoting the exchange stiffness constant. The length scale $l_H$ characterizes the range over which perturbations in the magnetization decay.

\begin{figure}[tbp]
	\begin{center}
		\advance\leftskip-0.4cm
		\advance\rightskip0cm
		\includegraphics[width=0.6\textwidth, angle=270]{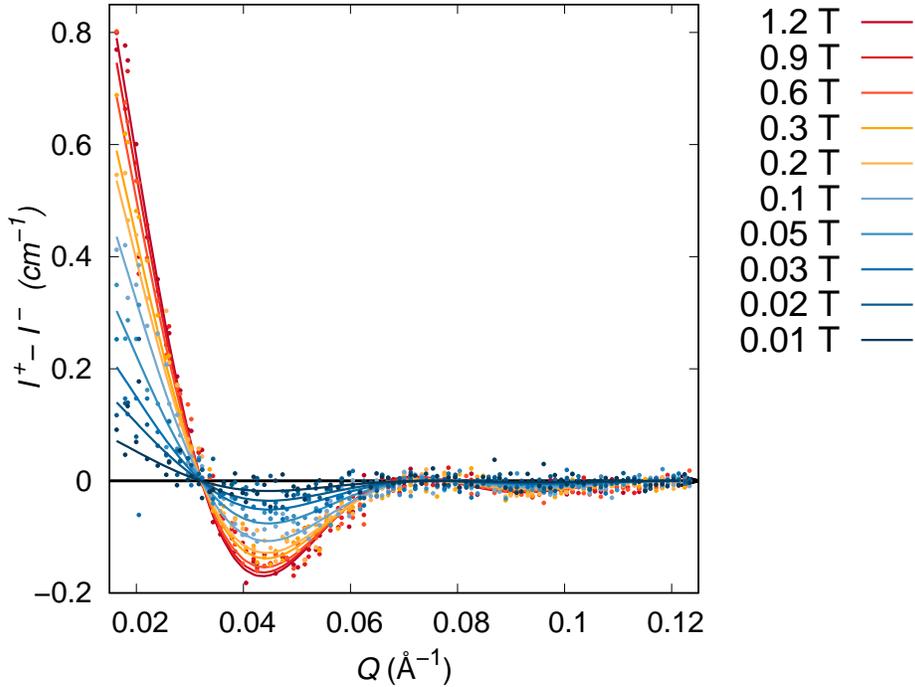}
		\caption{Fit of the SANSPOL cross term according to micromagnetic theory.}
		\label{fig:micromagn_fit}
	\end{center}
\end{figure}

Equation \eqref{eq:micromag} contains the Fourier coefficient $g_{H}(Q)$, which is independent of the applied magnetic field and contains information on the strength and the spatial structure of perturbing fields associated with the magnetic disorder anisotropy and fluctuations in magnetoelastic coupling. We assume a core-shell morphology for $g_H$, with a magnetic core having a reduced or even negligible perturbating disorder field and a surface shell with a drastically increased defect density giving rise to a random site perturbing field and hence misalignment of the magnetic moment from the magnetic easy axis of the particle.
The exchange interaction is not accessible from the fit due to the restricted Q range.

\section{Free energy calculation \label{sec:Edis}}
The field dependent Zeeman energy $E(H)$ of a nanoparticle in an external field is given by:
\begin{equation}\label{eq:2}
E(H) = - \mu\mu_0H\langle\cos\gamma\rangle = -\mu_0HV_\mathrm{mag}(H)M_\mathrm{z}(H),
\end{equation}
where $\mu_0$ is the permeability of free space and $\mu(H) = V_\mathrm{mag}(H)M_\mathrm{s}$ the integrated particle moment with $V_\mathrm{mag}(H) = \frac{4}{3}\pi r_\mathrm{mag}^3(H)$ the coherently magnetized volume at the magnetic field $H$. The longitudinal magnetization of the coherently magnetized particle core $M_\mathrm{z}(H) = M_\mathrm{s}\langle\cos\gamma(H)\rangle$ is directly accessible using polarized SANS (\eqref{eq:bH}). The Zeeman energy difference between the initial magnetized volume close to remanence and the increased magnetic volume for a specific applied magnetic field amounts to the energy required to align the disordered surface spins.

\begin{acknowledgments}
We thank Achim Rosch 
for fruitful discussion. We further acknowledge Jan Ducho\v{n} for HRTEM measurements and EDX maps at Research Centre Rez, Rez near Prague, Czech Republic, and Michael Smik for support during SAXS measurements at JCNS.
We gratefully acknowledge the financial support provided by JCNS to perform the neutron scattering measurements at the Heinz Maier-Leibnitz Zentrum (MLZ), Garching, Germany, and the provision of beamtime at the instrument D33 at the Institut Laue-Langevin, Grenoble, France.
The presented work was financially supported by the Ministry of Education, Youth and Sport Czech Republic (National Programme of Sustainability II), Project LQ1603 (Research for SUSEN).
Financial support from the German Research Foundation (DFG Grant DI 1788/2-1) as well as the Institutional Strategy of the University of Cologne within the German Excellence Initiative (Max Delbr\"{u}ck-Prize for Young Researchers 2017) are gratefully acknowledged.
In Memoriam D. Ni\v{z}\v{n}ansk\'{y}.
\end{acknowledgments}


%

\end{document}